\documentclass[12pt,english]{article}
\usepackage{babel}
\usepackage{amsfonts}
\date{}
\newcommand{\be}{\begin{equation}}
\newcommand{\ee}{\end{equation}}
\newcommand{\beq}{\begin{eqnarray}}
\newcommand{\eeq}{\end{eqnarray}}
\newcommand{\nn}{\nonumber}
\newcommand{\bi}{\bibitem}
\title{{\bf NJL interaction derived from QCD: vector and axial-vector mesons}
}

\author{B.A. ARBUZOV\\
{\it Skobeltsyn Institute of Nuclear Physics of
MSU,}\\ {\it 119992 Moscow, Russia}\\
{\it arbuzov@theory.sinp.msu.ru}\\
\\
 M.K. VOLKOV\\
{\it Joint Institute of Nuclear Research,}\\
{\it 141980 Dubna, Moscow Region, Russia}\\
{\it volkov@theor.jinr.ru}\\
\\
I.V. ZAITSEV\\
{\it Physics Department of MSU, 119992 Moscow, Russia}}

\begin{document}
\maketitle

\begin{quote}
In previous works effective non-local $SU(2)\times SU(2)$ NJL model was derived in 
the framework of the fundamental QCD. All the parameters of the model are expressed 
through QCD parameters: current light quark mass $m_0$ and average non-perturbative 
$\alpha_s$. The results for scalar and pseudo-scalar mesons are in satisfactory agreement to existing data. In the present work the same model without introduction 
of any additional parameters is applied for 
a description of masses and strong decay widths of $\rho$- and $a_1$-mesons. The results for both scalar and vector sectors agree with data with only one adjusted parameter 
$m_0$, with account of average  $\alpha_s \simeq 0.415$, which is obtained in 
a previous work as well.
\end{quote}

\section{Introduction}

For description of low-energy hadron physics phenomenological chiral quark Nambu -- 
Jona-Lasinio model \cite{NIL, EG, VE} (see also review \cite{VR} and references 
therein) is successfully applied for many years. Mass spectra of scalar, pseudo-scalar, 
vector and axial vector mesons and their low-energy interactions are satisfactorily described in this model \cite{VAN, ER, VECH}. In the simplest case of 
$SU(2)\times SU(2)$ chiral symmetry this model contains four arbitrary parameters: 
current mass $m_0$ of quark doublet $u,\,d$ (in approximation $m_u=m_d$), coupling constant $G_1$ of for-quark interaction of the scalar and pseudo-scalar quark currents in the chiral symmetric form, coupling constant  $G_2$ of four quark interaction of vector and axial-vector currents, ultra-violet cut-off parameter $\Lambda$.

 These four parameters allows to describe the pion, the $\sigma$-meson, vector mesons 
$\rho$ and $\omega$, axial-vector mesons $a_1$, masses and their main strong decays and weak pion decay constant $f_\pi$. However, only one of these parameters -- $m_0$ --  coincides with a parameter of fundamental QCD. So the very important problem is to express the rest parameters in terms of QCD parameters, e.g. $m_0$ and 
$\alpha_s$ (or $\Lambda_{QCD}$). But for a long time this problem was not solved in a sufficiently satisfactory form.
 
In recent works \cite{Namb2, Arvol} we have succeed in obtaining description of 
$SU(2)\times SU(2)$ NJL model using only QCD parameters. This becomes possible due to use of N.N. Bogoliubov compensation approach \cite{Bog2,Bog} (application of the approach to QFT problems is described in work \cite{Arb04}). As a result a non-local version of NJL model was obtained with uniquely defined form-factor. Thus ultra-violet divergences disappear, therefore there is no need of introduction of parameter 
$\Lambda$. Constants 
 $G_1$ and $G_2$ are expressed through $m_0$ and strong constant $\alpha_s$ in the non-perturbative region. 

Remind, that application of these results to the sector of scalar and pseudo-scalar mesons leads to satisfactory description of $\pi$ and $\sigma$ masses, constant of weak pion decay $f_\pi$ and of strong $\sigma \to \pi \pi $ decay. Emphasize, that only parameters $m_0$ and $\alpha_s$ were used.

It is worth noting, that independent estimate of average non-perturbative value 
$\alpha_s$ was obtained in work \cite{Arbplb} (see also \cite{arbary}). The same  Bogoliubov approach for a study of effective non-local three-gluon interaction results  in existence of the stable solution for the definite form of non-perturbative  contributions to running coupling  $\alpha_s(q^2)$. This corresponds to average value for the running coupling in the non-perturbative region 
$\alpha_s=0.415$ \cite{arbary}. Taking into account this result only one parameter 
$m_0$ remains in our disposal. Note, that results of works \cite{Arbplb}, \cite{arbary} 
lead to a consistent value of gluon condensate.

Here we use the version of of non-local NJL model obtained in  \cite{Arvol} with the same parameters $m_0$ and $\alpha_s$ for calculation of masses and decay widths of vector and axial-vector mesons $\rho$ and $a_1$. Remind that we introduce no new parameters at all. A special attention will be paid to value $\alpha_s=0.415$.

\section{Compensation equation for effective form-factor}

In the same way as in work~\cite{Arvol} we start from the standard Lagrangian of QCD with two light quarks and number of colours $N=3$
 \be
L\,=\,\sum_{k=1}^2\biggl(\frac{\imath}{2} \Bigl(\bar\psi_k\gamma_
\mu\partial_\mu\psi_k\,-
\,m_0\bar\psi_k\psi_k\,-\partial_\mu\bar\psi_k \gamma_\mu\psi
_k\,\biggr)+\,g_s\bar\psi_k\gamma_\mu t^a A_\mu^a\psi_k \biggr)\,-
\,\frac{1}{4}\,\biggl( F_{\mu\nu}^aF_{\mu\nu}^a\biggr);
\label{initial} \ee
In accordance to the approach, application of which to such
problems are described in details in work~\cite{Arb04}, we look
for a non-trivial solution of a compensation equation, which is
formulated on the basis of the Bogoliubov procedure {\bf add --
subtract}. Namely let us rewrite the initial
expression~(\ref{initial}) in the form
\beq & &L\,=\,\frac{\imath}{2} \Bigl(\bar\psi\gamma_
\mu\partial_\mu\psi-\partial_\mu\bar\psi\gamma_\mu\psi
\biggr)\,-\,\frac{1}{4}\,F_{0\,\mu\nu}^aF_{0\,\mu\nu}^a\,-
\,m_0\bar\psi\,\psi\,+ \,\frac{G_1}{2}\cdot\Bigl(
\bar\psi\tau^b\gamma_5\psi\,\bar\psi \tau^b\gamma_5
\psi\,-\nn\\
& &-\bar\psi\,\psi\,\bar\psi\,\psi\biggr)\,+\,\frac{G_2}
{2}\cdot\Bigl(\bar\psi\tau^b\gamma_\mu\psi\,\bar\psi
\tau^b\gamma_\mu\psi + \bar\psi \tau^b\gamma_5 \gamma_\mu \psi
\bar\psi \tau^b \gamma_5 \gamma_\mu \psi\biggr)\,+\nn\\&
&+\,\frac{G_3} {2}\cdot\Bigl(\bar\psi\gamma_\mu\psi\,\bar\psi
\gamma_\mu\psi + \bar\psi \gamma_5 \gamma_\mu \psi \bar\psi
\gamma_5 \gamma_\mu \psi\biggr)+\,g_s\,\bar\psi\gamma_\mu t^a
A_\mu^a\psi\,-\label{addsub}\\
& &-\,\frac{1}{4}\,\biggl( F_{\mu\nu}^aF_ {\mu\nu}^a -
F_{0\,\mu\nu}^aF_{0\,\mu\nu}^a\biggr)\,-\,\frac{G_1}{2}
\cdot\Bigl(\bar\psi \tau^b\gamma_5 \psi\,\bar\psi \tau^b
\gamma_5\psi -\bar\psi\,\psi\,\bar\psi\,\psi\biggr)\,-
\nn\\
& &-\,\frac{G_2}{2}\cdot\Bigl(\bar\psi \tau^b\gamma_\mu
\psi\,\bar\psi \tau^b\gamma_\mu \psi + \bar\psi \tau^b\gamma_5
\gamma_\mu \psi \bar\psi \tau^b \gamma_5 \gamma_\mu
\psi\biggr)\,-\nn\\& &-\,\frac{G_3}
{2}\cdot\Bigl(\bar\psi\gamma_\mu\psi\,\bar\psi \gamma_\mu\psi +
\bar\psi \gamma_5 \gamma_\mu \psi \bar\psi \gamma_5 \gamma_\mu
\psi\biggr).\label{intas} \eeq

Here $\psi$ is isotopic doublet, colour summation is performed inside each 
spinor bi-linear combination,
$F_{0\,\mu\nu} = \partial_\mu A_\nu - \partial_\nu A_\mu$, and e.g. 
notion $G_1\cdot \bar\psi \psi \bar\psi \psi$
means non-local vertex in the momentum space
 \be
\imath\,(2\pi)^4\,G_1\,F_1(p1,p2,p3,p4)\, \delta(p1+p2+p3+p4)\,,
\label{vertex} \ee
where form-factor $F_1$is introduced, which depends on incoming momenta. 
The Lagrangian contains contribution of both $G_2$ and $G_3$ which are connected correspondingly to isovector and isoscalar terms. In the present work we consider compensation equation only for isovector four-fermion terms.

Now we consider the first two lines of the Lagrangian (\ref{addsub}) as new free 
Lagrangian $L_0$, and the two last ones as interaction Lagrangianа $L_{int}$. 
Then
compensation conditions (see again~\cite{Arb04}, \cite{Arvol} will
consist in demand of full connected four-fermion vertices,
following from Lagrangian $L_0$, to be zero. This demand
gives a set of non-linear equations for form-factors
$F_i$. These equations according to terminology of 
works~\cite{Bog2, Bog} are called {\bf compensation equations}.

In a study of these equations the
existence of a perturbative trivial solution (in our case
$G_i = 0$) is always evident, but a non-perturbative
non-trivial solution may also exist. In the present work as well as in 
previous one we look for an adequate
approach, the first non-perturbative approximation of
which describes the main features of the problem.
Improvement of a precision of results is to be achieved
by corrections to the initial first approximation.

We follow works \cite{Namb2}, \cite{Arvol} in definition of 
the approximation.\\
1) In compensation equations we restrict ourselves by
terms with loop numbers 0, 1, 2. \\
2) In compensation equations we perform a procedure of
linearization over form-factor, which leads to linear integral
equations. It means that in loop terms only one vertex contains the
form-factor, while other vertices are considered
to be point-like. \\
3) While evaluating diagrams with point-like vertices diverging
integrals appear. Bearing in mind that as a result of the study we
obtain form-factors decreasing at momentum infinity, we use an 
intermediate regularization by introducing UV cut-off $\Lambda$ in the 
diverging integrals. It will be shown that results do not depend on the 
value of this cut-off.\\
4) We use a special approximation for integrals, which is connected with 
transfer of a quark mass from its propagator to the lower limit of 
momentum integration. Effectively this leads to introduction 
of IR cut-off at the lower limit of
integration by Euclidean momentum squared $q^2$ at value $m^2$. 
To justify this prescription let us consider a typical integral 
to be encountered here and perform simple evaluations. Functions which 
we use here depend on variable of the form $\alpha\,q^2$, where $\alpha$ 
is a parameter having $1/m^2$ dimension.

5) We keep the first two terms of $1/N$ expansion in equations. 

6) In case of vector vertices there are two Lorentz structures and thus we 
have generally speaking two form-factors instead of one in~\cite{Arvol}. 
However the corresponding set of equations has no explicit solution 
similar to that of~\cite{Arvol} and so we proceed in the following way. 
In our approximation we impose simplified kinematic condition that
left-side legs of diagrams have momenta $p$ and $-p$, while 
right-side ones have zero momenta. Now in addition to terms proportional 
to $\gamma_\varrho\times\gamma_\varrho$, which we are interested in, 
terms of the form  $\hat{p}\times\hat{p}$ may be also present, Supposing, 
that the presence of a form-factor connected with the last structure 
gives small corrections we shall transform the initial equation (in diagram 
form see Fig. 1) to the scalar one contracting it with projector of the form 
\be
\frac{1}{12}\,(\gamma_\varrho -
\frac{\hat{p}\,p_\varrho}{p^{2}}).
\ee
In the process of the study we have considered also equations obtained with use 
of projectors of more general form, namely
\be
\frac{1}{4\,(4-d)}\, (\gamma_\varrho -
d\,\frac{\hat{p}\,p_\varrho}{p^{2}}); 
\ee 
It becomes clear, that for values $d$ between 1 and 2 the corresponding solutions
lead to spread of physical values under interest in the range of $5\,-\,7\%$, 
that corresponds accuracy of the method as a whole. So we take the formulated projection procedure as a component of the first approximation. 

Now the demand of compensation of full connected four-fermion vertices 
proportional to $\it{G_2}$ multiplied by the vector form-factor leads us to the following equation (see Fig. 1) 
 \beq 
& &\ G_2\,F \left(
p^{2}\right)+
  \frac{{\it G_2}^{2}}{ {\pi }^{2}}\, \left( {\frac {65}{72}}\,{p}^{2}-{\frac
{7}{12 }}\,{p}^{2}\ln  \left( {\frac {{p}^{2}}{{\Lambda}^{2}}}
\right) -\frac{5}{4}\, {\Lambda}^{2} \right)+\nn\\
& &{\frac {{\it G_3}\,{\it G_2}}{{2}\, {\pi }^{2}}}\, \left( -
{\frac {43}{72}}\,{p}^{2}+{\frac {5}{12}}\,{p}^{2}\ln
 \left( {\frac {{p}^{2}}{{\Lambda}^{2}}} \right) +{\frac {3}{4}}\,{ \Lambda}^{2}
 \right) +\nn\\
& &{\frac {{\it G_2}^{2}\,N}{ {32\,\pi }^{6}}}\int
_{m_{{0}}}^{\infty }\!F \left( {k}^{2}
 \right)  \left( {\it G_2}^{2}\,N{\Lambda}^{2}-4\,{\pi }^{2} \right) {d^{4}k}+\,\frac{{{\it G_1}}^{2}}{{\pi }}
 \left( \frac{11}{288}\,{p}^{2}-\frac{1}{16}\,{\Lambda}^{2}-\frac{1}{48}\,{p}^{2}\ln  \left( {\frac {{p
}^{2}}{{\Lambda}^{2}}} \right)  \right) +\nn\\
& &+\frac{{{\it G_2}}^{3}N}{2{\pi }^{6}}\, \Biggl({\frac {7}{36}}
\int _{m_{{0}}}^{\infty }\! \, \left( \,{\frac {2{{\it
(kp)}}^{2}}{{p}^{2}}}+{k}^{2} \right)
 \left( p-k \right) ^{2}\ln  \left( {\frac { \left( p-k \right) ^{2}}{
{\Lambda}^{2}}} \right) \frac{F
\left( k^{2}\right){d^{4}k}}{(k^{2})^{2}}+\nn\\
& &+\int _{m_{{0}}}^{\infty }\! \left( -{ \frac {31}{108}}\,{\frac
{{{\it (kp)}}^{2}}{{p}^{2}}}-{\frac { 109}{864}}\,{k}^{2} \right)
\left( p-k \right) ^{2} \frac{F \left(
k^{2}\right){d^{4}k}}{(k^{2})^{2}}  +\int _{m_{{0}}}^{\infty
}\!\Biggl(\frac{1}{18}\,{k}^{2} {p}^{2}\ln \left( {\frac {
 \left( p-k \right) ^{2}}{{\Lambda}^{2}}} \right)+\nn\\
& &+\,\frac{3}{16}\left( \,{\frac {2{{\it
(kp)}}^{2}}{{p}^{2}}}+{k}^{2} \right){ \Lambda}^{2}}{ {
 }
+{(-{\frac {5}{432}})\left( -{{\it (kp)}}^{2}+3\,{k}^{2}{p}^{2}
\right) \Biggr) \frac{F
\left( k^{2}\right){d^{4}k}}{(k^{2})^{2}}}+\nn\\
& &+\int _{m_{{0}}}^{\infty }\!\Biggl( -\frac{1}{48}\,{\frac {
\left( {{\it (kp)}}^{2}-{k}^{2}{p}^{2} \right) {m_0}^{4}}{
 \left( p-k \right) ^{4}}}-{\frac {1}{96}}\,{\frac {{m_0}^{4} \left( 7\,
{k}^{2}{p}^{2}+8\,{{\it (kp)}}^{2} \right) }{ \left( p-k \right)
^{2}{p} ^{2}}}\Biggr)\frac{F
\left( k^{2}\right)d^{4}k}{(k^2)^2}\Bigr)+\nn\\
& &+ \frac{{{\it G_2}{\it G_3}}^{2}N}{2{\pi }^{6}}\, \Biggl( \int
_{m_{{0}}}^{\infty }\! \, \left( \,\frac{55}{576}{\frac {{{\it
(kp)}}^{2}}{{p}^{2}}}+\frac{17}{2304}{k}^{2} \right)
 \left( p-k \right) ^{2}\ln  \left( {\frac { \left( p-k \right) ^{2}}{
{\Lambda}^{2}}} \right) \frac{F
\left( k^{2}\right){d^{4}k}}{(k^{2})^{2}}+\nn\\
& &+\int _{m_{{0}}}^{\infty }\! \left( { \frac {17}{216}}\,{\frac
{{{\it (kp)}}^{2}}{{p}^{2}}}+{\frac { 1009}{13824}}\,{k}^{2}
\right) \left( p-k \right) ^{2} \frac{F \left(
k^{2}\right){d^{4}k}}{(k^{2})^{2}}  +\int _{m_{{0}}}^{\infty
}\!\Biggl(\frac{11}{384}\,{k}^{2} {p}^{2}\ln \left( {\frac {
 \left( p-k \right) ^{2}}{{\Lambda}^{2}}} \right)+\nn\\
& &-\,\frac{49}{768}\left( \,{\frac {2{{\it
(kp)}}^{2}}{{p}^{2}}}+{k}^{2} \right){ \Lambda}^{2}}{ {
 }
+{{\frac {5}{6912}}\left( 31{{\it (kp)}}^{2}-33\,{k}^{2}{p}^{2}
\right) \Biggr) \frac{F
\left( k^{2}\right){d^{4}k}}{(k^{2})^{2}}}+\nn\\
& &+\int _{m_{{0}}}^{\infty }\!\Biggl( \frac{1}{288}\,{\frac {
\left( {{\it (kp)}}^{2}-{k}^{2}{p}^{2} \right) {m_0}^{4}}{
 \left( p-k \right) ^{4}}}+{\frac {1}{384}}\,{\frac {{m_0}^{4} \left( 7\,
{k}^{2}{p}^{2}+8\,{{\it (kp)}}^{2} \right) }{ \left( p-k \right)
^{2}{p} ^{2}}}\Biggr)\frac{F \left(
k^{2}\right)d^{4}k}{(k^2)^2}\Bigr)+\nn\\
& &+ \frac{{{\it G_2}}{\it G_1}^{2}N}{2{\pi }^{6}}\, \Biggl(
-{\frac {1}{288}}\int _{m_{{0}}}^{\infty }\! \Biggl(\,\left(
\,{\frac {4 {{\it (kp)}}^{2}}{{p}^{2}}}-{k}^{2} \right) \,
 \ln  \left( {\frac { \left( p-k \right) ^{2}}{{\Lambda}^{2}}}
 \right) -\nn\\
& &-{\frac {1}{1728}}\,{\left( \,{\frac {32{{\it
(kp)}}^{2}}{{p}^{2}}}+{k}^{2} \right)} \Biggr)  \left( p-k \right)
^{2} \frac{F
\left( k^{2}\right){d^{4}k}}{(k^{2})^{2}}+\nn\\
& &+\int _{m_{{0}}}^{\infty }\!\Biggl(-{\frac {1}{864}}\, \left(
-5+6\,\ln  \left( {\frac {
 \left( p-k \right) ^{2}}{{\Lambda}^{2}}} \right)  \right)  \left( {k}
^{2}{p}^{2}-{{\it (kp)}}^{2} \right)
+\nn\\
& &+\,\frac{1}{96}\left( \,{\frac {2{{\it
(kp)}}^{2}}{{p}^{2}}}+{k}^{2} \right){ \Lambda}^{2}}{ {
 }
{\Biggr) \frac{F
\left( k^{2}\right){d^{4}k}}{(k^{2})^{2}}}+\nn\\
& &+\int _{m_{{0}}}^{\infty }\!\Biggl(-{\frac {1}{192}}\,{\frac
{{m_0}^{4} \left( 5\,{k}^{2}{p}^{2}-2\,{{\it kp}}^{2} \right) }{
\left( p-k \right) ^{2}{p}^{2}}}+1/48\,{\frac {{m_0} ^{4} \left(
{k}^{2}{p}^{2}-{{\it kp}}^{2} \right) }{ \left( p-k
 \right) ^{4}}}\Biggr)\frac{F \left(
k^{2}\right)d^{4}k}{(k^2)^2} \Biggr) =0
 \eeq 

Here a conversion to Euclidean momentum space is performed, at one-loop 
level terms proportional to $N$ and $1$ are taken into account and for two 
loops  respectively $N^{2}$ and $N$. The lower limit of integration is defined 
by current quark mass $m_0$ corresponding to value $u_0=1.925\,10^{-8}$, 
which is obtained in the course of consideration of scalar form-factor
(see~\cite{Namb2, Arvol}). We also use relation  
\be 
G_1=\frac{6}{13}\,G_2; \label{G_12} 
\ee
which is derived in the same works. After dividing by $G_2$ that correspond to our intention to find a non-trivial solution we integrate by angular variables of 
four-dimensional space we have 
\beq 
& &F \left( x
\right)+ \frac{N}{{\pi }^{4}}\, \Biggl(\Biggl(\left({\frac
{1}{64}}\,{{\it G_2}}^{2}-{\frac {1 }{256}}\,{{\it
G_3}}^{2}-{\frac {1 }{128}}\,{ {{{\it G_1}}^{2}}} \right)\int
_{m_0}^{x}\!  {\frac
{1}{x}} \quad F \left( y \right) {dy}+\nn\\
& &+\left( {\frac {1}{96}}\,{{ \it G_2}}^{2}+{\frac {1}{96}}\,{\it
G_3}\,{\it G_2}+{\frac {1}{384}}\,{{{{\it G_1}}^{2}}{}}\right)\int
_{m_0}^{x}\! {\frac {y}{{x}^{2}}}
 F \left( y \right) {dy} \Bigr){m_0}^{4}+ \nn\\
& &+\left( {\frac {13}{96}}\,{{\it G_2}}^{2}-{\frac {5}{
96}}\,{\it G_3}\,{\it G_2}+{\frac {1}{192}}\,{{\it
G_1}}^{2}+{\frac {1}{ 384}}\,{{\it G_3}}^{2} \right)\ln  \left( x \right) x \int _{m_0}^{x}\!\ F \left( y \right) {dy}+\nn\\
& &+\left( {\frac {1}{8}}\,{{ \it G_2}}^{2}-{\frac {7}{96}}\,{\it
G_3}\,{\it G_2} \right)\ln  \left( x \right) \int _{m_0}^{x}\!y\,
F
\left( y \right) {dy} + \nn\\
& &+\left( {\frac {7}{32}}\,{{\it G_2}}^{2}-{\frac
{29}{288}}\,{\it G_3}\, {\it G_2}+{\frac {1}{384}}\,{{\it
G_3}}^{2}+{\frac {1}{192}}\,{{\it G_1}} ^{2} \right) \int
_{m_0}^{x}\!y\, F \left( y
\right) {dy}+ \nn\\
& &+ {\frac {1}{x}} \left( -{\frac {13}{1152}}\,{\it G_3}\,{\it
G_2}+{\frac {1}{128}}\,{{\it G_2}}^{2}-{\frac {1}{1536}}\,{{\it
G_3}}^{2}-{\frac {1} {768}}\,{{\it G_1}}^{2} \right) \int
_{m_0}^{x}\!\,{y}^{2} F \left( y \right) {dy}+ \nn\\
& &+ \left( {\frac {1}{ 11520}}\,{{\it G_3}}^{2}-{\frac
{1}{1920}}\,{\it G_3}\,{\it G_2}+{\frac { 7}{2880}}\,{{\it
G_2}}^{2}
 +{\frac {{1}}{5760}}\,\it G_1^{2}\right){\frac {{1}}{{x}^{2}}}
\int _{m_0}^{x}\!\,{y}^{3} F \left( y \right) {dy} + \nn\\
& &+\Biggl(\left( {\frac {7}{ 384}}\,{\it G_1^{2}}+  {\frac
{5}{192}}\,{{ \it G_2}}^{2}+{\frac {1}{24}}\,{\it G_3}\,{\it
G_2}+{\frac {1}{256}}\,{{\it G_3}}^{2}
  \right) x\int _{x}^{\infty }\!    {\frac {1}{
{y}^{2}}}\,F \left( y \right) {dy}+ \nn\\
& &+\left( -{\frac {3}{128}}\,{\it G_1}^{2}-{\frac
{1}{128}}\,{{\it G_3}}^{2}-\frac{1}{32}\,{\it G_3}\,{\it G_2}
 \right) \int _{x}^{\infty }\!    {\frac {1}{
{y}}}\,F \left( y \right) {dy} \Bigr) {m_0}^{4}+\nn\\
& &+  \left( -{\frac {1}{3840}}\,{{\it G_1}}^{2}  - {\frac
{1}{7680}}\,{{\it G_3}}^{2}+{ \frac {1}{640}}\,{{\it
G_2}}^{2}-{\frac {13}{5760}}\,{\it G_3}\,{\it G_2}
  \right) {x}^{3} \int _{x}^{\infty }\!    {\frac {1}{
{y}^{2}}}\,F \left( y \right) {dy}+\nn\\
& &+  \left( {\frac {5}{1152}}\,{\it G_3} \,{\it G_2}+{\frac
{1}{384}}\,{{\it G_1}}^{2}+{\frac {1}{64}}\,{{\it G_2}
}^{2}+{\frac {1}{768}}\,{{\it G_3}}^{2} \right) {x}^{2}\int
_{x}^{\infty }\!
{\frac {1}{ {y}}}\,F \left( y \right) {dy}+\nn\\
& &+
   \left( {\frac {13}{96}}\,{{\it G_2}}^{2}-{\frac {5}{96}}\,{
\it G_3}\,{\it G_2}+{\frac {1}{192}}\,{{\it G_1}}^{2}+{\frac
{1}{384}}\,{ {\it G_3}}^{2} \right)\,x \int _{x}^{\infty }\! \,\ln
\left( y \right)\,F \left( y \right) {dy} +\nn\\
& &+\left({\frac {61}{288}}\,{{\it G_2}}^{2}+{\frac
{1}{1152}}\,{{\it G_3}}^{2}-{\frac {11}{96}}\,{\it G_3} \,{\it
G_2}+{\frac {1}{576}}\,{{\it G_1}}^{2} \right)\, x\int
_{x}^{\infty
}\! \,F \left( y \right) {dy}+ \nn\\
& &+\left(  {\frac {1}{8}}\,{{ \it G_2}}^{2}-{\frac {7}{96}}\,{\it
G_3}\,{\it G_2} \right)\int _{x}^{\infty }\! \ln \left( y
 \right) y \, F \left( y \right) {dy}\Bigr)+\nn\\
& &+\frac{N}{{\pi }^{4}}\, \Biggl(\Biggl(\left( {\frac
{1}{192}}\,{\it G_3}\,{\it G_2}-{\frac {1}{32}}\,{{ \it G_2}}^{2}
\right) x\int _{m_0}^{\infty }\! {\frac {1}{ {y}^{2}}}\,F\,{dy}+\nn\\
& &+ \left( {\frac {1}{64}}\,{\it G_3}\,{ \it G_2}-{\frac
{3}{32}}\,{{\it
G_2}}^{2}\right) \int _{m_0}^{\infty }\! {\frac {1}{ {y}}}\,F\,{dy}  \Bigr) {m_0}^{ 4}+ \label{EXY}\\
& &+ \left( -{\frac {13}{96}}\,{{\it G_2}}^{2}-{\frac {1}{192}}
\,{{\it G_1}}^{2}+{\frac {5}{96}}\,{\it G_3}\,{\it G_2}-{\frac
{1}{384}} \,{{\it G_3}}^{2} \right) \ln \left( {\Lambda}^{2}
\right)\,x\int _{m_0}^{\infty }\! \,F\,{dy} +\nn\\
& &+\left(-{\frac { 119}{576}}\,{{\it G_2}}^{2}-{\frac
{11}{2304}}\,{{\it G_3}}^{2}+{\frac { 43}{576}}\,{\it G_3}\,{\it
G_2}-{\frac {11}{1152}}\,{{\it G_1}}^{2}
 \right) x\int _{m_0}^{\infty }\! \,F\,{dy}+ \nn\\
& &+ \left( {\frac {7}{96}}\,{\it G_3}\,{\it G_2}-{\frac {1}{
8}}\,{{ \it G_2}}^{2} \right) \ln  \left( {\Lambda}^{2}
\right)\int
_{m_0}^{\infty }\!y \,F\,{dy} + \nn\\
& &+\left( {\frac {53}{ 576}}\,{\it G_3}\,{\it G_2}-{\frac
{1}{192}}\,{{\it G_1}}^{2}-{\frac {1}{ 384}}\,{{\it
G_3}}^{2}-{\frac {19}{96}}\,{{\it G_2}}^{2} \right) \int
_{m_0}^{\infty }\!y
\,F\,{dy}+\nn\\
& &+
 \left( -{\frac {3}{32}}\,{\it G_3}\,{\it G_2}+{\frac {1}{128}}\,{{\it
G_3}}^{2}+{\frac {9}{32}}\,{{\it G_2}}^{2}+{\frac {1}{64}}\,{{\it
G_1}}^{ 2} \right) {\Lambda}^{2}  \int _{m_0}^{\infty }\!F
\left( y \right) {dy}
 \Bigr)\Bigr)+\nn\\
& &+{\frac { {\it G_2} \,N \left( {\it
G_2}\,N{\Lambda}^{2}-4\,{\pi }^{2} \right) }{16\,{\pi }^{4} }}\int
_{m_0}^{\infty }\!\,F \left( y \right){dy}+{\frac {{\it G_2}}{{\pi
}^{2}}}\, \left( {\frac {65}{72}}\,x-{\frac {7}{12}}\,x\ln
 \left( {\frac {x}{{\Lambda}^{2}}} \right) -{\frac {5}{4}}\,{\Lambda}^{2}  \right) +
\nn\\
& &{\frac {{\it G_3}}{{2}\, {\pi }^{2}}}\, \left( - {\frac
{43}{72}}\,x+{\frac {5}{12}}\,x\ln
 \left( {\frac {x}{{\Lambda}^{2}}} \right) +{\frac {3}{4}}\,{ \Lambda}^{2}
  \right)+\nn\\
& &+{\frac {{{\it G_1 }}^{2}  }{{{288\,\it G_2}\,\pi }^{2}}}\left(
11\,x-18\,{\Lambda}^{2}-6\,x\ln
 \left( {\frac {x}{{\Lambda}^{2}}} \right)  \right)=0\,.\nn
 \eeq

In view of looking for solutions of Eq.~(\ref{EXY}) we apply the differential operator
\beq 
& &\frac{d^3}{dx^3}\,x\,\frac{d^2}{dx^2}x\, \frac{d^3}{dx^3}x^{2}\,
\nn ,\eeq
to this equation. As a result we obtain a differential equation, 
which with account of the following substitution
\be 
z=\beta\,x^{2}\,,\; \beta =
\frac{1}{2^{6}}\,\frac{{N}\,\it{G_2}\left(
 12\it{G_2}-7\it{G_3}\right)}{24\,{\pi}^{4}}\,;
\ee 
reduces to the following form
\beq 
& &\biggl(z\,\frac{d}{dz}-b_1\biggr)\biggl(z\,\frac
{d}{dz}-b_2\biggr)\,\biggl(z\,\frac{d}{dz}-b_3\biggr)\biggl(z\,\frac{d}{dz}-
b_4\biggr)\biggl(z\,\frac{d}{dz}-b_5\biggr) \biggl
(z\,\frac{d}{dz}-b_6\biggr)\times\nn\\
&
&\times\biggl(z\,\frac{d}{dz}-b_7\biggr)\biggl(z\,\frac{d}{dz}-b_8\biggr)
 F(z) =z\,\biggl
(z\,\frac{d}{dz}-a_1+1\biggr)\biggl (z\,\frac{d}{dz}-a_2+1\biggr)
F(z)\,;\label{DE}
\eeq
i.e. it is Meijer equation of the eighth order. Solutions of the equation are represented in terms of the Meijer functions~\cite{be} with parameters  
$b_i,\, a_i$, which we can calculate provided 
$\it{G_3}$ is defined. We can naturally admit $\it{G_3}=\it{G_2}$ following 
rules of NJL model (see \cite{VECH}). In what follows we present confirmation 
of this assumption. In this case we have
\beq 
& &b_1 :=1.5\,;\quad
                            b_2 = 1\,;\quad
                           b_3 = 0.499991384\,;\quad
                       b_4 = .500008866\,;\quad\nn\\
& &
                    b_5 = -1.45597130 \cdot10^{-7}\,;\;
                            b_6 = 0\,;\; b_7 = -0.50000003\,;
\; b_8 = -1.0000001\,;\nn\\
& &\quad a_1 := -0.3944464\,;\quad
                       a_2 := 1.9013991\,.\label{ab}
\eeq 
Values of parameters are calculated with account of value $m_0$.

To obtain a solution of the integral equation we choose four linearly independent solutions of Eq.~(\ref{DE}) decreasing at infinity and form the following linear 
combination with coefficients $C_i$ 
\beq 
& &F(z)\,=\,C_1\,G_{2\,8}^{5\,1}\biggl(\,z\,|^{a_1,\,a_2}_{b5, b4, b3, b2, b1, b8, b7, b6}\biggr)+\,C_2\,G_{2\,8}^{5\,1}\biggl(z\,|^{a_1,\,a_2}_{b6, b5, b3, b2, b1, b8, b7, b4}\,\biggr)+\nn\\
&
&+\,C_3\,G_{2\,8}^{5\,1}\biggl(z\,|^{a_1,\,a_2}_{b7, b4, b3, b2, b1, b8, b6, b5}\biggr)+\,C_4\,G_{2\,8}^{7\,1}\biggl(z\,|^{a_1,\,a_2}_{b8, b6, b5, b4,
b3, b2, b1, b7}\Bigr)
\,.
\eeq 
Coefficients $C_i$ are fixed by boundary conditions, which are obtained in the same way as in work~\cite{Arb04}
\beq 
& & {3}\, \left( {\frac {13}{96}}\,{{\it
G_2}}^{2}+{\frac {1}{192}}\,{{\it G_1}}^{2} +{\frac
{1}{384}}\,{{\it G_3}}^{2}-{\frac {5}{96}}\,{{\it G_2}}{{\it
G_3}}\right) {\frac {1}{ {\pi }^{4}\sqrt {\beta}}}\int
_{m_0^{2}}^\infty \! F
 \left( y \right) {dy}-\nn\\
& &-{\frac {7}{12}}\,{\frac {{{\it G_2}}}{{ \pi
}^{2}}}-\frac{1}{48}\,{\frac {{{\it G_1}}^{2}}{{\pi }^{2}{\it
G_2}}}+\frac{5}{24}\,{\frac
{{{\it G_3}}}{{\pi }^{2}}}=0\,;\label{BC}\\
& &\int _{m_0^{2}}^{\infty }\! y\,F
 \left( y \right) {dy}=0\,;\quad
\int _{m_0^{2}}^{\infty }\!
y^{2}\,F
 \left( y \right) {dy=0}\,;\quad\int _{m_0^{2}}^{\infty }\!
y^{3}\,F
 \left( y \right) {dy}=0\,.\nn
\eeq
As a result we have
\beq 
& & C_1 :=0.3330348455\,;\quad
                   C_2 := 6.254973002 \cdot10^{-8}\,;\\\nn
& &
                   C_3 :=3.452159489 \cdot10^{-8}\,;\quad
                   C_4 := 2.105889777 \cdot10^{-15}\,.
                   \eeq

Unlike of scalar case~\cite{Arvol} we here do not force  
the form-factor value at lower integration limit to be unity. 
Using this condition one might try to define ratio of 
 $\it{G_2}$ and $\it{G_3}$. However assuming equality of these 
constants we avoid solution of additional complicated transcendental 
equation, but we acquire a criterion of self-consistency of our 
approach as a whole, because calculations show, that changing this ratio 
in reasonable range we have satisfactory results for values of the form-factor at 
the normalization point. 
In our case we have $ F(u_0)=0.96094 $
and so we consider our assumption to be justified with reasonable accuracy.
Admissible are values of ratio $\frac{\it G_2}{\it G_3}=\chi$ 
from $1$ up to $1.2$ as well. For the last value $ F(u_0)=1.098993576$.
As a matter of fact to fix the ratio one should consider also 
equation for isoscalar vector terms. However this leads to 
a considerable complication of the procedure and so here we 
only noting, that preliminary estimates show that just for range    
 $\chi=1\, -\, 1.2$ values of isoscalar vector form-factor 
differs from unity not more than by $10\%$. So admitting $\chi=1$ we
formulate the ground approximation bearing in mind necessity of 
further corrections.

\section{Wave function of vector states}

We have the non-trivial solution of the compensation equation 
and thus four-fermion terms are excluded from \textbf{free} 
Lagrangian. There is of course no compensation in \textbf{interaction} Lagrangian, which contains these terms with opposite sign. So we can study 
a problem of bound states with account of this four-fermion interaction. 
Bethe -- Salpeter equation for vector case in the same approximation 
as above (see Fig. 2) has the following form . Remind that the first approximation corresponds to zero-mass states (in this approximation 
there is the same equation for vector and axial-vector).
\beq 
& &\Psi(y)=  \frac{N}{{\pi }^{4}}\, \Biggl({m}^{4}\Biggl(\left({\frac
{3}{256}}\,{{\it G_2}}^{2}-{\frac {{ {{{\it G_1}}^{2}}} }{128}} \right)\int
_{m^2}^{x}\!  {\frac
{1}{x}} \quad \Psi \left( y \right) {dy}+\nn\\
& &+\left( {\frac {{{ \it G_2}}^{2}}{48}}+{\frac {{{{\it G_1}}^{2}}}{384}}\right)\int
_{m^2}^{x}\! {\frac {y}{{x}^{2}}}
 \Psi \left( y \right) {dy} \Bigr)+ \left( {\frac {23}{11520}}\,{{\it
G_2}}^{2}
 +{\frac {{{\it G_1^{2}}}}{5760}}\right)
\int _{m^2}^{x}\!{\frac {{y^3}}{{x}^{2}}} \Psi \left( y \right) {dy} + \nn\\
& &+\left( {\frac {11}{128}}\,{{\it G_2}}^{2}+{\frac {{{\it
G_1}}^{2}}{192}} \right)\ln  \left( x \right) x \int _{m^2}^{x}\!\ \Psi \left( y \right) {dy}+ {\frac {5}{96}}\,{{ \it G_2}}^{2}\ln  \left( x \right) \int _{m^2}^{x}\!y\,
\Psi
\left( y \right) {dy} + \nn\\
& &+\left( {\frac {139}{1152}}\,{{\it G_2}}^{2}+{\frac {{{\it G_1}} ^{2}}{192}}\, \right) \int
_{m^2}^{x}\!y\, \Psi \left( y
\right) {dy}- \left( {\frac {19}{4608}}\,{{\it G_2}}^{2}+{\frac {{{\it G_1}}^{2}} {768}} \right) \int
_{m^2}^{x}\!\frac{{y}^{2}}{x} \Psi(y) {dy}+ \nn\\
& &+\Biggl(\left( {\frac {7 \it {G_1^{2}}}{ 384}} +  {\frac
{55}{768}}\,{{ \it G_2}}^{2}
  \right) \int _{x}^{\infty }\!    {\frac {x}{
{y}^{2}}} \Psi \left( y \right) {dy}- \left( {\frac {3 {\it G_1}^{2}}{128}}\,+{\frac
{5 {{\it G_2}}^{2}}{128}}\,
 \right) \int _{x}^{\infty }\!    {\frac {\Psi \left( y \right)}{
{y}}}  {dy} \Bigr) {m}^{4}-\nn\\
& &-  \left( {\frac {{{\it G_1}}^{2}}{3840}}  +{\frac
{19 {{\it G_2}}^{2}}{23040}}\,
  \right)  \int _{x}^{\infty }\!    {\frac {{x}^{3}}{
{y}^{2}}}\,\Psi \left( y \right) {dy}+  \left( {\frac
{{{\it G_1}}^{2}}{384}}\,+{\frac {49}{2304}}\,{{\it G_2}
}^{2} \right) \int
_{x}^{\infty }\!
{\frac {{x}^{2}}{ {y}}}\,\Psi \left( y \right) {dy}+\nn\\
& &+
   \left( {\frac {11{{\it G_2}}^{2}}{128}}+{\frac {{{\it G_1}}^{2}}{192}} \right)x \int _{x}^{\infty } \ln
\left( y \right) \Psi \left( y \right) {dy} +\left({\frac {113}{1152}} {{\it G_2}}^{2}+{\frac {{{\it G_1}}^{2}}{576}}  
\right)\, x\int
_{x}^{\infty
}\! \,\Psi \left( y \right) {dy}+ \nn\\
& &+ {\frac {5}{96}}\,{{ \it G_2}}^{2} \int _{x}^{\infty }\! \ln \left( y
 \right) y \, \Psi \left( y \right) {dy}\Bigr)+\frac{N}{{\pi }^{4}}\, \Biggl(\Biggl( -{\frac {5}{192}}\,{{ \it G_2}}^{2}
\int _{m^2}^{\infty }\! {\frac {x}{ {y}^{2}}}\,\Psi\,{dy}+\label{BS}\\
& &-{\frac
{5}{64}}\,{{\it
G_2}}^{2} \int _{m^2}^{\infty }\! {\frac {\Psi}{ {y}}}\,{dy}  \Bigr) {m}^{ 4}- \left( {\frac {11}{128}}\,{{\it G_2}}^{2}+{\frac {{{\it G_1}}^{2}}{192}}
 \right) \ln \left( {\Lambda}^{2}
\right)\,x\int _{m^2}^{\infty }\! \,\Psi\,{dy} +\nn\\
& &+\left(-{\frac {35}{256}}\,{{\it G_2}}^{2}-{\frac {11}{1152}}\,{{\it G_1}}^{2}
 \right) x\int _{m^2}^{\infty }\! \,\Psi\,{dy}-{\frac {5}{96}}\,{{ \it G_2}}^{2}  \ln  \left( {\Lambda}^{2}
\right)\int
_{m^2}^{\infty }\!y \,\Psi\,{dy} + \nn\\
& &+\left( -{\frac
{{{\it G_1}}^{2}}{192}}-{\frac {125}{1152}}\,{{\it G_2}}^{2} \right) \int
_{m^2}^{\infty }\!y
\,\Psi\,{dy}+
 \left( {\frac {25}{128}}\,{{\it G_2}}^{2}+{\frac {{{\it
G_1}}^{ 2}}{64}} \right) {\Lambda}^{2} \Bigr) \int _{m^2}^{\infty
}\!\Psi \left( y \right) {dy}
 \Bigr)\Bigr)+\nn\\
& &\left( \alpha_s-\frac{3}{8}{\frac {{g_{{v}}}^{2}}{\pi }}
\right)\Biggl(\frac{1}{9{\pi }} \int _{m^{2}}^{\,x }\!
 \Psi
 \left( y \right)
 \left( \frac{15}{x}+{\frac {2\,y}{{x}^{2}}} \right) {dy}+\frac{1}{9\pi} \int
_{x}^{\infty }\! \Psi
 \left( y \right)\left(\frac{12}{y}+\frac{5\,x}{y^{2}} \right){dy}\Biggr)\,.\nn 
\eeq

Here besides the same kernel as in Eq.(\ref{EXY}) we take into account also one-gluon exchange and one-meson exchange with corresponding 
constants  $\alpha_s$ and $g_v^2/4 \pi$. Note that contributions of (pseudo-)scalar 
mesons here cancel. In Eq.~(\ref{BS}) enters constituent mass $m$ instead of current 
mass in~Eq.(\ref{EXY}). For parameter $m$ we use results of previous work~\cite{Arvol} 
where it was obtained from stability condition for the effective potential. This 
procedure allows to define $m$ corresponding to value of $\alpha_s$. In the same way as 
in~\cite{Arvol} we take values of $u=\beta\,m^{4}$, which correspond to values of 
$\alpha_s$ in  
the range under study. We perform calculations for 
$u\,=\, 0.00015,\, 0.00030,\, 0.00045$. Values of $\alpha_s$, 
which are slightly corrected in comparison with that of work~\cite{Arvol} are presented in the summarizing table. 

Differential equation now is the following 
\beq 
& &\biggl(z\,\frac{d}{dz}-b_1\biggr)\biggl(z\,\frac
{d}{dz}-b_2\biggr)\,\biggl(z\,\frac{d}{dz}-b_3\biggr)\biggl(z\,\frac{d}{dz}-
b_4\biggr)\biggl(z\,\frac{d}{dz}-b_5\biggr) \biggl
(z\,\frac{d}{dz}-b_6\biggr)\times\nn\\
& &\times\biggl(z\,\frac{d}{dz}-b_7\biggr)\biggl(z\,\frac{d}{dz}-b_8\biggr)
 \Psi(z) =-z\,\biggl
(z\,\frac{d}{dz}-a_1+1\biggr)\biggl (z\,\frac{d}{dz}-a_2+1\biggr)
\Psi(z)\,;\label{DEBS} 
\eeq 
where 
\beq 
& & z=\beta x^{2}\,,\quad \beta =
\frac{1}{2^{6}}\,\frac{{N}\,\it{G_2}\left(
 12\it{G_2}-7\it{G_3}\right)}{24\,{\pi}^{4}}\,; 
\quad \xi=\frac{{\it G_1}}{{\it G_2}}\,;\nn\\
& & a_1={\frac {1}{80}}\,{\frac {59\,{\xi}^{2}+6-\sqrt
{8281\,{\xi}^{4}+708\,{ \xi}^{2}+36}}{{\xi}^{2}}}\,; \nn\\
& & a_2={\frac {1}{80}}\,{\frac {59\,{\xi}^{2}+6+\sqrt
{8281\,{\xi}^{4}+708\,{ \xi}^{2}+36}}{{\xi}^{2}}}\,;
\eeq 
and coefficients $b_i$ are roots of the following equation
( ${\it G_3}={\it G_2}$)
\beq 
& &
\Biggl(-{\frac {435}{5408}}\,{\frac {N{m}^{4}{{\it
G_2}}^{2}{b}^{4}}{{\pi }^{4 }}}+{\frac {11539}{21632}}\,{\frac
{N{m}^{4}{{\it G_2}}^{2}{b}^{5}}{{ \pi }^{4}}}-\nn\\
& & -{\frac {28195}{21632}}\,{\frac {N{m}^{4}{{\it
G_2}}^{2}{b}^{3}}{{\pi }^{4}}}+{\frac {1217}{2704}}\,{ \frac
{N{m}^{4}{{\it G_2}}^{2}{b}^{2}}{{\pi }^{4}}}+{\frac {2859}{5408}
}\,{\frac {N{m}^{4}{{\it G_2}}^{2}b}{{\pi }^{4}}}
 \Biggr) + \\
& & +\left( 16\,{\frac {b}{ \pi }}-{\frac {38}{3}}\,{\frac
{{b}^{5}}{\pi }}+{\frac {20}{3}}\,{ \frac {{b}^{4}}{\pi }}+{\frac
{110}{3}}\,{\frac {{b}^{3}}{\pi }}+\frac{8}{3}\, {\frac {{b}^{6}}{\pi
}}-{\frac {148}{3}}\,{\frac {{b}^{2}}{\pi }}
 \right)  \left( \alpha_{{s}}-\frac{3}{8}\,{\frac {{g_{{v}}}^{2}}{\pi }}
 \right)-\nn\\
& &  -2\,{b}^{6}-16\,{b}^{3}+12\,{b}^{2}+20\,{b}^{5}+{b}^{8}-4\,{b
}^{7}-11\,{b}^{4}=0\,.\nn 
\eeq
Solution of Eq.~(\ref{DEBS}) decreasing at infinity has the following general form
\beq 
& & \Psi(z)\,=\,C_1\,G_{2\,8}^{4\,1}\biggl(\,z\,|^{a_1,\,a_2}_{b1, b2, b3, b5, b4, b8, b7, b6}\biggr)+\,C_2\,G_{2\,8}^{4\,1}\biggl(z\,|^{a_1,\,a_2}_{b1, b2, b3, b4, b5, b6, b7, b8}\,\biggr)+\nn\\
& &+\,C_3\,G_{2\,8}^{4\,1}\biggl(z\,|^{a_1,\,a_2}_{b1, b2, b5, b6, b3, b4, b7, b8}\biggr)+C_4\,G_{2\,8}^{6\,1}\biggl(z\,|^{a_1,\,a_2}_{b1,
b2, b3, b4, b5, b7, b6, b8}\biggr)+\label{SolBS}\\
& &+\,C_5\,G_{2\,8}^{6\,1}\biggl(z\,|^{a_1,\,a_2}_{b1, b2, b3, b5,
b6, b8, b7, b4}\biggr)\,.\nn
\eeq 
For $u=0.00030$ values of parameters $b_i$ read
\beq 
& & b_1=1.5; \quad b_2=1\,;\qquad
b_3=0.5\,;\qquad b_4=0.763584407\,;\label{bi}\\
& &b_5=0.19323742\,; \; b_6=0\,; \; b_7=-0.7956972342\,; \; b_8=-1.161124600\,.
\nn 
\eeq
Parameters $a_i$ are the same as before~(\ref{ab}). 
Coefficients $C_i$ are defined from the boundary conditions
 \beq
& &   \Psi\left(
m^{2} \right)=1\,;\quad \int _{m^{2}}^{\infty }\! \Psi
 \left( y \right) {dy}=0\,;\quad \int _{m^{2}}^{\infty }\! y\,\Psi
 \left( y \right) {dy}=0\,;\nn\\
& &\int _{m^{2}}^{\infty }\! y^{2}\,\Psi
 \left( y \right) {dy=0}\,;\quad \int _{m^{2}}^{\infty }\! y^{3}\,\Psi
 \left( y \right) {dy}=0\,;
\eeq
and value $g_{{v}}$ is given by the iterative procedure being defined 
by normalization condition in one-loop approximation
\be 
\frac{N\,g_v^{2}}{12\,\pi^{2}}\int _{\tilde{u}}^{\infty }\!\,\frac{\Psi
 \left( z \right)^{2} F
 \left( z \right)}{z}{dz}=1\,;\quad \tilde{u}=\frac{\beta}{\beta_0}\,u\,; \quad 
\beta_0\,=\,\frac{(G_1^2+6\,G_1 G_2)\,N}{16\,\pi^4}\,.\label{gv}
\ee
Here we introduce into the integral form-factor $F\left( z \right)$ which was obtained in the previous section in view to take into account decreasing of interaction 
for increasing momentum variable. 

Ratio 
\be
\frac{\beta}{\beta_0}=\frac{845}{754}
\ee 
gives coefficients for transitions to variable $z \sim {p}^{4}$ respectfully for vector and scalar sectors. Expression for $\beta_0$ is obtained in works~\cite{Namb2}, \cite{Arvol}

With parameters  $b_i$~(\ref{bi}) we have
\beq & & C_1=1.7465; \quad C_2=0.021266; \quad C_3=0.00107221; \quad C_4=
0.00142116; \nn\\ 
& & C_5=-0.0000525341\,; \quad g_v=5.00\,.\label{Ci}
\eeq 

\section{Results and discussion}

Now we proceed to calculation of observable parameters. Initial estimate of $\rho$-meson mass is given by expression~\cite{VECH}  
\be 
M_0 =\frac{{g_v}}{\sqrt{{\it G_2}}}\,;
\ee
where $\it{G_2}$ is defined from relation~(\ref{G_12})  
and $\it{G_1}$ is calculated according the method of work~\cite{Arvol} for 
chosen value 
$u$. Then we introduce one-loop correction to mass squared, which is given by 
expression
\beq\Delta(M_0^{2})=-\frac{3\,g_v^{2}}{8\,\pi^{2}\,\sqrt{\beta}}\int
_{\tilde{u}}^{\infty }\!\,\frac{\Psi
 \left( z \right)^{2}\,F
 \left( z \right)}{\sqrt{z}}{dz} \,; 
\eeq
so that 
\be
M_\rho=\sqrt{M_0^{2}+ \Delta(M_0^{2})}\,.
\ee 
For values of parameters presented above we have
\be 
M_0=974\,MeV\,; \quad
\quad \Delta(M_0^{2})=-325908\,MeV^2\,; \quad \quad M_\rho=789 MeV\,.
\ee
We also estimate  $a_1$meson mass according to well known relation~\cite{VECH} 
\be 
M_{a_1}^{2}={M_\rho}^{2}
+ 6\,m^{2}\,.\label{ma1}
\ee
Coupling constant  $g_{\rho \rightarrow 2\,\pi}$ of $\rho$-decay to two $\pi$-mesons  
we find with triangle diagram according to the following relation
\beq 
g_{\rho \rightarrow 2\,\pi}=
{g_s}^{2}\,g_v \, \frac{3}{4\,\pi^2} \int _{u}^{\infty
}\!\,\frac{\Psi_s
 \left( z \right)^{2}\,\Psi\left( \frac{\beta}{\beta_0}\,z \,\right)\,F
 \left( z \right)}{z}{dz}\,;\label{V2S}
\eeq
where $\Psi_s
 \left( z \right)$ is Bethe Salpeter wave function for scalar states 
and $g_s$ is scalar meson coupling according to definition in work~\cite{Arvol}.

The width of $\rho$ is the following
\be 
\Gamma_\rho={\frac {g_{\rho\mapsto 2\,\pi }^{2} \left( {M_{{\rho}}}^{2}-
4\,{m_{{\pi}}}^{2} \right) ^{3/2}}{24\, \pi {M_{{\rho}}}^{2}}}\,.
\ee

For $a_1$-meson width we consider two channels: $a_1 \to \rho \,\pi$ and 
$a_1 \to \sigma \,\pi$. The vertex for the first decay has the following form (we omit obvious isotopic factor $\epsilon_{abc}$)
\be
 V_{\mu \nu}(a_1 \to \rho \,\pi)\,=\,A_0\,g_{\mu \nu}+A_2\,p_\nu\,q_\mu\,;\label{a1rho}
\ee
$$
A_0\,=\,-\,\frac{N g_v^2 g_s m}{\pi^2}\,\int_{m^2}^\infty \frac{\Psi(y)^2\,\Psi_s(y)}{y}\,dy\,;\quad A_2\,=\,\frac{N g_v^2 g_s m}{2\,\pi^2}\,\int_{m^2}^\infty \frac{\Psi(y)^2\,\Psi_s(y)}{y^2}\,dy\,;
$$
where $p,\,\mu$ and $q,\,\nu$ are respectfully momentum and Lorentz index for 
$a_1$-meson and $\rho$-meson, $m$ is constituent quark mass.

The second decay is described by the following vertex (isotopic factor $\delta_{ab}$)
\beq 
& & V_\mu (a_1 \to \sigma \,\pi)\,=\,g_{a_1 \to \sigma \pi}\,(q-k)_\mu\,;\label{a1sig}\\
& & g_{a_1 \to \sigma \pi}\,=\,\frac{N g_v g_s^2 }{2\,\pi^2}\,\int_{m^2}^\infty \frac{\Psi(y)\,\Psi_s(y)^2}{y}\,dy\,;\nn
\eeq
where $k$ and $q$ are respectfully momentum of $\pi$ and $\sigma$ and $\mu$ is Lorentz index of $a_1$.
Corresponding partial widths read
\beq
& & \Gamma(a_1 \to \rho \pi)\,=\,\frac{M_{a1}^2-M_\rho^2}{24 \pi M_{a1}^3}\Biggl(
A_0^2\biggl(2+\frac{(M_{a1}^2+M_\rho^2)}{4 M_{a1}^2 M_\rho^2}\biggr)+\nn\\
& &+A_0 A_2\frac
{(M_{a1}^2+M_\rho^2)(M_{a1}^2-M_\rho^2)^2}{4 M_{a1}^2 M_\rho^2}+A_2^2\frac
{(M_{a1}^2-M_\rho^2)^4}{16 M_{a1}^2 M_\rho^2}\Biggr)\,;\label{a1ro}\\
& & \Gamma(a_1 \to \sigma \pi)\,=\,\frac{g_{a_1 \to \sigma \pi}^2(M_{a1}^2-M_\sigma^2)^3}{48 \pi M_{a1}^5}\,;\quad \Gamma_{a_1}\,=\,\Gamma(a_1 \to \rho \pi)+
\Gamma(a_1 \to \sigma \pi)\,.\nn
\eeq 
Here we assume $m_\pi^2 << M_{a_1,\rho,\sigma}^2$.
 
Using all these expressions we calculate observable quantities for vector and axial-vector mesons. Note, that in calculation of widths of decays we substitute calculated masses of corresponding mesons. Results are presented in Table. We present there 
set of calculated parameters in dependence on average non-perturbative running coupling $\alpha_s$ in range 0.29 -- 0.48. We normalize our calculations by the most precise 
parameter $f_\pi$. All other numbers in the Table are calculated. We take each column of the Table as a set of the corresponding parameters calculated starting from 
presented there parameters $\alpha_s$ and $m_0$. Following the general ideology of 
our approach we consider the last column with $\alpha_s=0.415$ as the final result of 
our work, bearing in mind, that this value of average $\alpha_s$ is obtained in 
work~\cite{arbary}. In addition to values of parameters presented in this column we 
remind parameters of scalar sector, which are obtained in the course of performing 
of previous work~\cite{Arvol}. For the same $\alpha_s=0.415$ we have: 
\be
m_\pi\,=\,134\,MeV\,;\quad <\bar q\, q>\,=\,- (230\,MeV)^3\,;\quad 
m_\sigma \,=\,480\,MeV\,;\quad  \Gamma_\sigma\,=\, 560\,MeV\,.\label{scal}
\ee
Thus the set of parameters seems to be in satisfactory agreement with data. 
The only parameter differing from corresponding experimental value by more than 
12\% is the mass of $a_1$. As a matter of fact a low value for $M_{a_1}$ is inherent 
to other NJL calculations (see, e.g.~\cite{VECH}). Presumably in this case there are some additional contributions to be taken into account.

The results being obtained here and in previous work~\cite{Arvol} demonstrate that Bogoliubov compensation method leads to a reasonable form of effective non-local 
four-quark interaction  of the NJL type. As a result this method allows to describe the light meson masses and probabilities of their main decays. Emphasize, that advantages 
of this approach are absence of ultra-violet divergences in quark loop diagrams, inherent in the usual NJL model, and the presence of only one arbitrary parameter, 
namely current quark mass $m_0$. Note that in spite of this parameter being somewhat larger than its standard value we nevertheless obtain reasonable value for constituent quark mass $m \simeq 260 MeV$~\cite{Arvol}. Note, that value of $m_0$ is defined by compensation 
equation for scalar form-factor in works~\cite{Namb2, Arvol}. Estimates show that 
presence of correction terms in this equation (e.g. the next terms of $1/N$ expansion) 
changes value of $m_0$ significantly, while the observable parameters change 
only slightly. A development of these considerations is the problem for future studies. 

This method in future studies may also be applied to description of electro-weak properties of mesons (e.g. pion form-factor, pion polarizability), of $\pi - \pi$ scattering lengths etc. It is important problem to expand the approach for chiral 
$U(3)\times U(3)$ symmetry with inclusion of the strange quark $s$.

\newpage
\begin{center}
{\bf Table}\\
\bigskip
\bigskip
\begin{tabular}{|c|c|c|c|c|c|}
\hline
$u$ & 0.00015 & 0.00030 & 0.00045 & 0.00032 & exp/phen\\
\hline
$\alpha_s$ & 0.287 & 0.404 & 0.483 & 0.415 & 0.415~\cite{arbary} \\
\hline
$f_\pi\,MeV$ & 93 & 93 & 93 & 93 & input\\
\hline
$g_s$& 2.66 & 2.84 & 2.93 & 2.86 & -- \\
\hline
$m_0\,MeV$ & 21.9 & 21.6 & 21.2 & 21.5 & 5 -- 10 \\
\hline 
$m\,MeV$ &247  & 264 & 271 & 265 &  
$\simeq 300$ \\
\hline
$G_1^{-1/2}\,MeV$ &320&  287 & 267 & 283 &-- \\
\hline
$g_v$ & 4.30 & 5.00 & 5.52 & 5.11 & --\\
\hline 
$M_\rho\,MeV$ & 713 & 785 & 830 & 791
& 771.1$\pm$0.9\\
\hline
$g_{\rho\,\pi\,\pi}$ & 4.29 & 4.41 & 4.36 & 4.44 & 4.26\\
\hline $\Gamma_\rho\,MeV\;$ & 136 & 166 &
175 & 170 & 149.2 $\pm$0.7\\
\hline
$M_{a_1}\,MeV$ & 935 & 1017 & 1043 & 1018 & 1230 $\pm$40\\
\hline
$\Gamma_{a_1}\,MeV$ & 268 & 330 & 312 & 334 & 250 -- 600\\
\hline
$\Gamma(a_1 \to \sigma\,\pi)/\Gamma_{a_1}$ & 0.168 & 0.188 & 0.201 & 0.189 & $0.188 \pm 0.043\,$\cite{asn} \\
\hline
\end{tabular}
\end{center}

\newpage
\begin{center}
{\bf Figure captions}
\end{center}
\bigskip
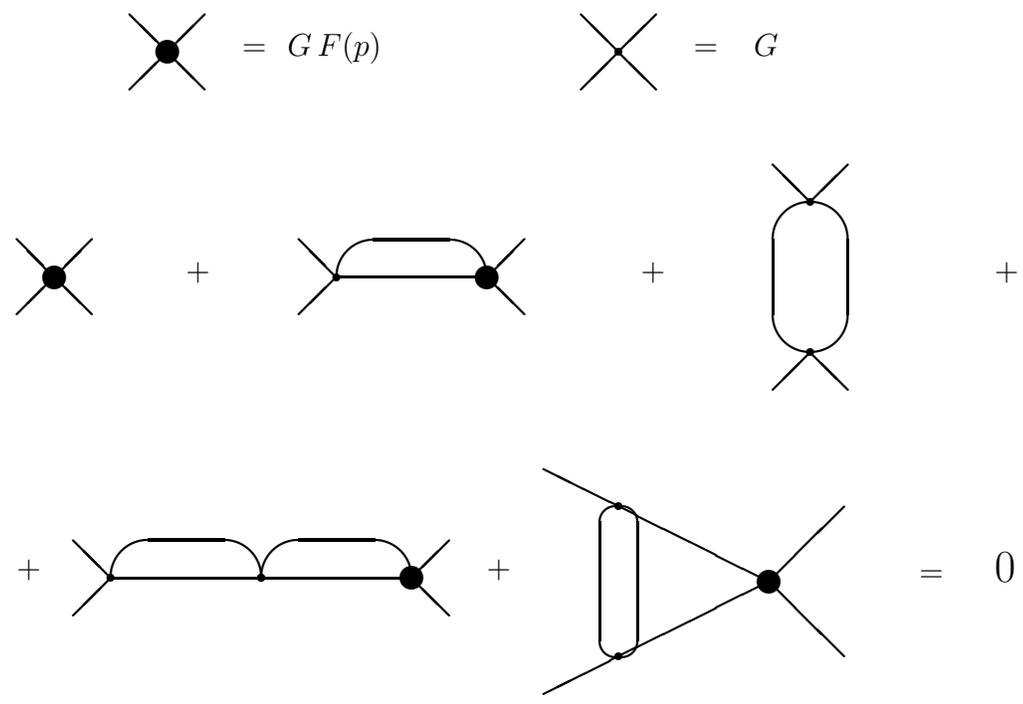
Fig. 1. Diagram representation of the compensation equation.\\
\\
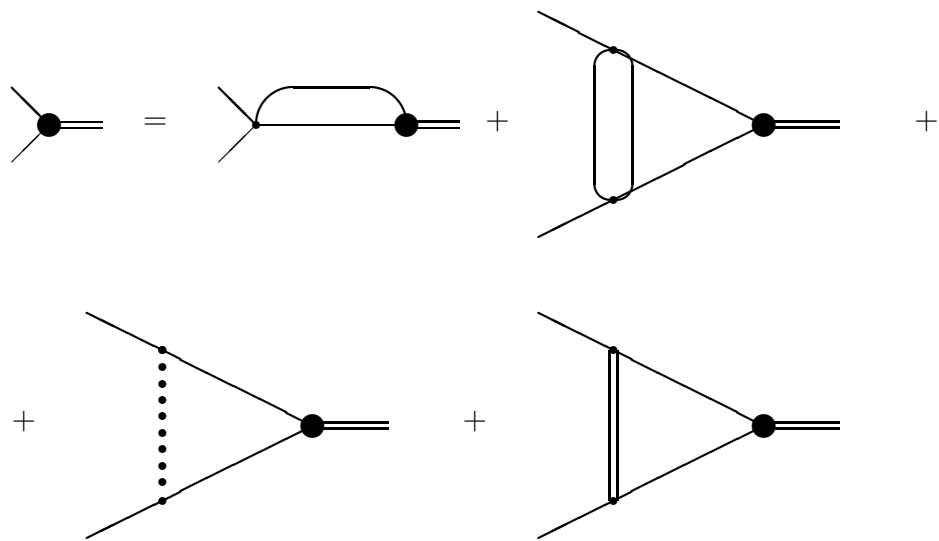
Fig. 2. Diagram representation of Bethe-Salpeter equation for vector bound state.
\newpage 
\begin{picture}(160,85)
{\thicklines \put(20,80.5){\line(-1,1){5}}
\put(20,80.5){\line(1,1){5}} \put(20,80.5){\circle*{3}}
\put(20,80.5){\line(-1,-1){5}} \put(20,80.5){\line(1,-1){5}}
\put(30,80){=} \put(36,80){$G\,F(p)$} {\thicklines
\put(80,80.5){\line(-1,1){5}} \put(80,80.5){\line(1,1){5}}
\put(80,80.5){\circle*{1}} \put(80,80.5){\line(-1,-1){5}}
\put(80,80.5){\line(1,-1){5}}} \put(90,80){=} \put(98,80){$G$}

{\thicklines \put(5,50.5){\line(-1,1){5}}
\put(5,50.5){\line(1,1){5}} \put(5,50.5){\circle*{3}}}
\put(5,50.5){\line(-1,-1){5}} \put(5,50.5){\line(1,-1){5}}

\put(22.5,50){+} {\thicklines \put(42.5,50.5){\line(-1,1){5}}
\put(52.5,50.5) {\oval(20,10)[t]}
\put(62.5,50.5){\line(1,1){5}}\put(42.5,50.5) {\circle*{1}}
\put(62.5,50.5){\circle*{3}}} \put(42.5,50.5){\line(-1,-1){5}}
\put(62.5,50.5) {\line(1,-1){5}} \put(42.5,50.5){\line(1,0){20}}
\put(83,50){+}

{\thicklines \put(105.5,60.5){\line(-1,1){5}}
\put(105.5,60.5){\line(1,1){5}} \put(105.5,50.5){\oval(10,20)}
\put(105.5,60.5){\circle*{1}} \put(105.5,40.5){\circle*{1}}}
\put(105.5,40.5){\line(-1,-1){5}} \put(105.5,40.5){\line(1,-1){5}}
\put(130,50){+} \put(0,10.5){+} {\thicklines
\put(12.5,10.5){\line(-1,1){5}} \put(22.5,10.5) {\oval(20,10)[t]}
\put(12.5,10.5) {\circle*{1}} \put(32.5,10.5){\circle*{1}}}
\put(12.5,10.5){\line(-1,-1){5}} \put(12.5,10.5){\line(1,0){20}}
{\thicklines
 \put(42.5,10.5)
{\oval(20,10)[t]} \put(52.5,10.5){\line(1,1){5}}\put(32.5,10.5)
{\circle*{1}} \put(52.5,10.5){\circle*{3}}}
 \put(52.5,10.5)
{\line(1,-1){5}} \put(32.5,10.5){\line(1,0){20}}
\put(62.5,10.5){+} {\thicklines \put(100,10){\line(-2,1){30}}
\put(100,10){\line(-2,-1){30}} \put(80,10){\oval(5,20)}
\put(80,20){\circle*{1}} \put(80,0){\circle*{1}}
\put(100,10){\line(1,1){10}} \put(100,10){\line(1,-1){10}}
\put(100,10){\circle*{3}}}} \put(120,10){=} \put(130,10){{\Large
0}}
\end{picture}
\ \
\bigskip
\bigskip
\bigskip
\begin{center}
Fig. 1.
\end{center}
\newpage
\begin{picture}(160,55)

{\thicklines \put(5,40.5){\line(-1,1){5}}
\put(5,40.5){\circle*{3}}} \put(5,40.5){\line(-1,-1){5}}
\put(5,40.9){\line(1,0){7}} \put(5,40.1){\line(1,0){7}}
\put(17.5,40){=} {\thicklines \put(32.5,40.5){\line(-1,1){5}}
\put(42.5,40.5){\oval(20,10)[t]} \put(52.5,40.9){\line(1,0){7}}
\put(32.5,40.5) {\circle*{1}} \put(52.5,40.5){\circle*{3}}}
\put(32.5,40.5){\line(-1,-1){5}} \put(52.5,40.1){\line(1,0){7}}
\put(32.5,40.5){\line(1,0){20}} \put(63,40){+} {\thicklines
\put(100,40.5){\line(-2,1){30}} \put(100,40.5){\line(-2,-1){30}}
\put(80,40.5){\oval(5,20)} \put(80,50.5){\circle*{1}}
\put(80,30.5){\circle*{1}} \put(100,40.9){\line(1,0){10}}
\put(100,40.1){\line(1,0){10}} \put(100,40.5){\circle*{3}}}}
\put(120,40){+} \put(0,0){+} {\thicklines
\put(40,0.5){\line(-2,1){30}} \put(40,0.5){\line(-2,-1){30}}
\multiput(20,10.5)(0,-2.2){9}%
{\circle*{1}} \put(20,10.5){\circle*{1}}
\put(20,-9.5){\circle*{1}} \put(40,0.9){\line(1,0){10}}
\put(40,0.1){\line(1,0){10}} \put(40,0.5){\circle*{3}}}
\put(60,0){+} {\thicklines \put(100,0.5){\line(-2,1){30}}
\put(100,0.5){\line(-2,-1){30}} \put(80.5,-9.5){\line(0,1){20}}
\put(79.5,-9.5){\line(0,1){20}} \put(80,10.5){\circle*{1}}
\put(80,-9.5){\circle*{1}} \put(100,0.9){\line(1,0){10}}
\put(100,0.1){\line(1,0){10}} \put(100,0.5){\circle*{3}}
\end{picture}
\bigskip
\bigskip
\bigskip
\bigskip
\bigskip
\bigskip
\bigskip
\bigskip
\bigskip
\bigskip
\bigskip

\begin{center}
Fig. 2.
\end{center}

\end{document}